\documentclass[pra,twocolumn,superscriptaddress,floatfix,showpacs]{revtex4}
\usepackage{amsmath,epsf,dcolumn}

\newcommand{\be}{\begin{equation}}
\newcommand{\ee}{\end{equation}}
\newcommand{\bea}{\begin{eqnarray}}
\newcommand{\eea}{\end{eqnarray}}
\begin{document}

\title{Application of self-consistent $\alpha$ method to improve the performance of model exchange potentials}


\author{Valentin V.~Karasiev}
\email{vkarasev@qtp.ufl.edu}
\author{Eduardo V.~Lude\~na}
\affiliation{Centro de Qu\'{\i}mica, Instituto Venezolano de Investigaciones
Cient\'{\i}ficas, Apartado 21827,\\ Caracas 1020-A, Venezuela}
\author{Art\"em E.~Masunov}
\affiliation{NanoScience Technology Center, Department of Chemistry and
Department of Physics, University of Central Florida, 12424 Research Parkway, Orlando, Florida 32826}

\date{17 October 2009} 

\begin{abstract}
Self interaction error remains an impotrant problem in density functional theory.
A number of approximations to exact exchange aimed to correct for this error while 
retainining computational efficiency had been suggested recently.
We present a critical comparison between model exchange potentials generated 
through the application of the asymptotically-adjusted self-consistent
$\alpha$, AASC$\alpha$, method and BJ effective exchange potential advanced in [A.D. Becke and
E.R. Johnson, J. Chem. Phys. 124, 221101 (2006)] and
[V.N. Staroverov, J. Chem. Phys. 129, 134103 (2008)].  In particular
we discuss their compliance with coordinate-scaling, virial and
functional derivative conditions.  We discuss the application of the
AASC$\alpha$ method to generate the AA-BJ potential. A numerical
comparison is carried out through the implementation of a
fully-numerical diatomic molecule code yielding molecular virial
energies and ionization potentials approximated by the energies of the
HOMO orbitals. It is shown that some of the shortcomings of these
model potentials, such as the non-compliance with the Levy-Perdew
virial relation, may be eliminated by multiplying the response term by
an orbital-dependent functional $\alpha$, which can be simplified to a constant determined during
the self-consistent procedure (self-consistent $\alpha$).

\end{abstract}

\maketitle

\section{Introduction}

An important problem arising in the application of the Kohn-Sham
equations is that of the construction of the local exact exchange potential.
In principle, this potential can be exactly calculated following the
procedure outlined in the OEP method.
\cite{Sharp-Horton-1953,Talman-Shadwick-1976,Krieger-Li-Iafrate-1992,Gorling-Levy-1994,Gorling-1999,Ivanov-Hirata-Bartlett-1999,Yang-Wu-2002,YAW-2004,Mori-Sanchez-et-al-2005}
However, in practice, there are a number of difficult numerical
impediments that bar the way to the realization of this exact
approach\cite{Hirata-et-al-2001,Heaton-Burgess-et-al-2007,Gorling-et-al-2008,Pino-et-al-2009}
in addition to some deeper problems stemming from general instabilities
of Kohn-Sham potentials in finite dimensional
subspaces.\cite{Pino-et-al-2009}

For this reason, particularly in recent years, much attention has
been devoted to the development of alternative local-exchange
potentials which are simple to apply but which at the same time yield
sufficiently accurate
results.\cite{Krieger-Li-Iafrate-1992,Gritsenko-et-al-1995,Sala-Goerling.2001,Gritsenko-Baerends-2001,Harbola-Sen-2002,Holas-Cinal-2005,Umezawa-2006,Staroverov-Scuseria-2006,Becke-Johnson-2006,Izmaylov-et-al-2007,Staroverov08}

A few years ago, we introduced such an approach, which we called the
\lq\lq asymptotically-adjusted self-consistent $\alpha$\rq\rq\,\,
(AA-SC$\alpha$) method.  Although the detailed theoretical
justification for this method is presented in
Ref. \cite{Karasiev-AA-02}, we comment here on two of its
characteristics.

The first is that in the AA-SC$\alpha$ method the
following model potential is postulated:
\begin{equation}
v^{\rm AASC\alpha}_{\rm x}({\bf r})=v_{\rm S}({\bf r})+
\alpha_{\rm x}[\{\psi_i\}]\tilde v^{0}_{\rm resp}({\bf r})
\label{x1}
\end{equation}
where $v_{\rm S}({\bf r})$ is the local Slater potential and where
$\tilde v^0_{\rm resp}({\bf r})=v^0_{\rm x}({\bf r})-2\epsilon^0_{\rm
  x}([\rho];{\bf r})$ is a local response potential related to an
approximate (and arbitrary) exchange functional $E^0_{\rm
  x}[\rho]=\int d^3{\bf r}\rho({\bf r})\epsilon^0_{\rm x}([\rho];{\bf
  r}) $ yielding the local exchange potential $v^0_{\rm x}({\bf
  r})=\delta E^0_{\rm x}[\rho]/\delta\rho $; (however, more generally,
the response term $\tilde v^0_{\rm resp}$ also may be modelled).  In
Eq. (\ref{x1}), $\alpha_{\rm x}[\{\psi_i\}]$ is the functional
\begin{equation}
\alpha_{\rm x}[\{\psi_i\}]
={{E_{\rm x}[\{\psi_i\}]-E_{\rm xLP}\big[v_{\rm S}([\rho]),\rho\big]}
\over {E_{\rm xLP}\big[\tilde v^0_{\rm resp}([\rho]),\rho\big]}}
\label{x2}
\end{equation}
where $E_{\rm x}[\{\psi_i\}]$ is the exact orbital expression for the exchange energy,
and $E_{\rm xLP}\big[v([\rho];{\bf r}),\rho({\bf r})\big]
=\int d^3{\bf r}v({\bf r})[3\rho({\bf r})+ {\bf r}\cdot\nabla\rho({\bf r})])$ is the Levy-Perdew expression for
the exchange energy functional corresponding to the potential $v({\bf
  r})$ \cite{LevyPerdew85,ZhaoLevy93}.  
The local potential given by Eq. (\ref{x1}) is called \lq\lq
asymptotically adjusted\rq\rq\,\, because while the first term
guarantees the correct asymptotic behavior of $-1/r$ for large $r$ the
second term does not contribute in the asymptotic region.

The second characteristic follows from Eq. (\ref{x2}):
\begin{eqnarray}
E_{\rm x}[\{\psi_i\}]&=&E_{\rm xLP}\big[v_{\rm S}([\rho]),\rho\big]
+\alpha_{\rm x}[\{\psi_i\}]E_{\rm xLP}\big[\tilde v^0_{\rm resp}([\rho]),\rho\big]\nonumber\\
& \equiv & E^{\rm AASC\alpha}_{\rm x}[\{\psi_i\}]
\label{x3}
\end{eqnarray}
The variational derivative of this functional with respect to the
Kohn-Sham orbital yields:
\begin{equation}
{{\delta E^{\rm AASC\alpha}_{\rm x}[\{\psi_i\}]}\over{\delta \psi_j}}=
\bigg(v^{\rm AASC\alpha}_{\rm x}({\bf r})
+\Delta \widehat v_{\rm x}({\bf r})\bigg)\psi_j({\bf r})
\label{x4}
\end{equation}
where $\Delta \widehat v_{\rm x}({\bf r})\psi_j({\bf r})=[\widehat v_{{\rm x}j}({\bf
  r})-v_{\rm S}({\bf r})-\alpha_{\rm x}\tilde v^0_{\rm resp}({\bf r})]\psi_j({\bf r})$ is a
non-local correction to the $v^{\rm AASC\alpha}_{\rm x}({\bf r})$ potential.  It
has been shown \cite{Karasiev-AA-02} that the contribution of $\Delta
\widehat v_{\rm x}({\bf r})$ to the energy is $\Delta E_{\rm x}[\{\psi_i\}] = 0$
where $\Delta E_{\rm x}[\{\psi_i\}] = E_{\rm x}[\{\psi_i\}]-E_{\rm xLP}[v_{\rm S}]
-\alpha_{\rm x}[\{\psi_i\}]E_{\rm xLP}[\tilde v^0_{\rm resp}]$.  Hence, the
omission of $\Delta \widehat v_{\rm x}({\bf r})$ in the Kohn-Sham equation
only affects the quality of the converged orbitals but it does not
change the expression for the energy.

The AASC$\alpha$ method has been applied previously to improve the
potentials and energies of several DFT exchange functionals.  In
particular, we have analyzed the improvements brought about by this
method with respect to the LDA and PW91
functionals and proposed two models for $\tilde v^0_{\rm resp}$ term \cite{Karasiev-AA-02}.

However, quite recently a very simple potential denoted as the BJ
effective exchange potential has been proposed in a heuristic way by
Becke and Johnson\cite{Becke-Johnson-2006}. This potential contains a
Slater term plus a local response one.  This work has been extended by
Staroverov \cite{Staroverov08} who has advanced a family of model
potentials which have the form of a Slater potential plus a response
term which is modeled. In all these cases the energy is evaluated
using the exact expression $E_{\rm x}[\{\psi_i\}]$ for the exchange
functionals constructed from the $N$ occupied orbitals which have
self-consistently converged for the given model potential.
 
In the present work, due to the similarities between this family of
potentials and the AASC$\alpha$ one, we make a critical comparison
between these potentials. We also apply the AASC$\alpha$ method using
the model exchange functional associated with the BJ potential to
generate the AA-BJ one.  The systems chosen for the present comparison
are some selected diatomic molecules. We show that application of the
AASC$\alpha$ method does indeed bring improvements, albeit slight, on
both the BJ energies and ionization potentials. Also, we show that it
yields internuclear distances that are in excellent agreement with the
exact Kohn-Sham x-only results.

\section{Comparison with the BJ and related models}

There are three formal conditions that the exact optimized effective
potential (OEP) for exchange must
satisfy\cite{Ou-Yang.Levy.PRL.1990,Ou-Yang.Levy.PRA.44.1991}.  These
are: the variational derivative condition, the virial relation, and
the scaling requirement.

Quite clearly, the AASC$\alpha$ local potential $v^{\rm
AASC\alpha}_{\rm x}({\bf r})$ is an approximate one and, hence, it
does not satisfy all three of these conditions. As it is shown in
Eq. (\ref{x4}), the variational derivative of $E^{\rm AASC\alpha}_{\rm
  x}[\{\psi_i\}]$ with respect to $\psi_i$ yields the potential
$v^{\rm AASC\alpha}_{\rm x}({\bf r})+\Delta \widehat v_{\rm x}({\bf
  r})$. 
Let us emphasize, however, that $\Delta \widehat
v_{\rm x}({\bf r})$ does not contribute to the exchange energy, 
omission of this term in the Kohn-Sham equation makes the local potential 
$v^{\rm AASC\alpha}_{\rm x}$ the approximate one. 
With respect to the virial relation, it follows from the definition of
$v^{\rm AASC\alpha}_{\rm x}({\bf r})$ given by Eq. (\ref{x1}) 
that $v^{\rm AASC\alpha}_{\rm x}({\bf r})$ satisfies by construction
the Levy-Perdew virial condition.  Also, it is easy to show using the
fact that both $E_{\rm x}[\rho_{\lambda}]=\lambda E_{\rm x}[\rho]$ and
$E_{\rm xLP}\big[v([\rho_{\lambda}]),\rho_{\lambda}\big]
=\lambda E_{\rm xLP}\big[v([\rho]),\rho\big]$ that
$\alpha_{\rm x}[\{\psi_i\}]$, as defined by Eq. (\ref{x2}), is
invariant under coordinate scaling (does not depend on
$\lambda$). Then using $ E^{\rm AASC\alpha}_{\rm
  x}[\rho_{\lambda}]=E_{\rm xLP}\big[v_{\rm S}([\rho_{\lambda}]),\rho_{\lambda}\big]
+\alpha_{\rm x} E_{\rm xLP}\big[\tilde v^0_{\rm resp}([\rho_{\lambda}]),\rho_{\lambda}\big]$ and
the fact that $\delta E^{\rm AASC\alpha}_{\rm
  x}[\rho_{\lambda}]=\lambda \delta E^{\rm AASC\alpha}_{\rm x}[\rho]$,
and following the same arguments as in
Ref. \cite{Ou-Yang.Levy.PRA.44.1991}, it can be readily shown that
$v^{\rm AASC\alpha}_{\rm x}({\bf r})$ also satisfies the scaling
property.

In the case of the potentials introduced by Becke and Johnson
\cite{Becke-Johnson-2006} and Staroverov \cite{Staroverov08}, it is
clear that they satisfy the scaling requirement.  However, the fact
that these potentials do not satisfy the Levy-Perdew condition implies
that the virial relation in the Born-Oppenheimer approximation
\cite{Slater60} is not satisfied either (see Ref.  \cite{KL01} for
details). Moreover, as a consequence of this, \lq\lq there is no
unique choice of functional for the evaluation of the exchange energy"
for structure optimization or total energy comparison (see
Ref. \cite{Tran..Schwartz2007}, where the Becke-Johnson potential was
employed for band gap calculations in solids).  Also in the case of
the BJ and other model potentials discussed
in\cite{Becke-Johnson-2006,Staroverov08}, the third requirement from
\cite{Ou-Yang.Levy.PRL.1990,Ou-Yang.Levy.PRA.44.1991}, namely, that
the model potential must correspond to the functional derivative of
the model functional with respect to the density is not satisfied.

We denote as in Ref. \cite{Staroverov08} by $E_{\rm conv}$  
the total energy computed with the
exact exchange expression $E_{\rm x}[\{\psi_i\}]$ using the converged
orbitals and by $E_{\rm vir}$ the corresponding one which includes the
Levy-Perdew expression $E_{\rm xLP}[v_{\rm x},\rho]$ for
exchange. Since for the BJ and related potentials $E_{\rm x}[\{\psi_i\}]\ne
E_{\rm xLP}[v_{\rm x},\rho]$ the total energy values $E_{\rm conv}$
and $E_{\rm vir}$ are also different. The claims that \lq\lq $(E_{\rm
  vir}- E_{\rm conv})$ gives an indication of how close $v_{\rm
  x\sigma}$ is to the exact functional derivative"\cite{Staroverov08}
or that it serves as an indication of the accuracy of the calculation
itself (see Ref. \cite{Engel.Vosko.1993}) do not seem to hold
generally, as there are approximate KS exchange potentials (such as
the AASC$\alpha$ one, for example), which while differing from the
exact one, yield, nonetheless, $E_{\rm vir}= E_{\rm conv}$.

In the present article, in order to compare the results obtained
using the AASC$\alpha$ model with those coming from the BJ and related
potentials, we provide numerical values, in particular, for diatomic
molecules. In order to carry out this numerical comparison, the BJ
effective model potentials proposed in \cite{Becke-Johnson-2006} and
\cite{Staroverov08} were implemented in a fully-numerical diatomic
molecule code \cite{KLSS96} (due to its numerical instability stemming
from \lq\lq troublesome \rq\rq, terms, a fact that was corroborated in
our test calculations, the gradient-corrected model
\cite{Staroverov08} was not implemented).

For completeness, we include some results obtained in the context of
the GLLB model proposed in Ref.  \cite{Gritsenko-et-al-1995}, the
localized Hartree-Fock (LHF) model
\cite{Sala-Goerling.2001,Sala.Goerling.2003}, and the common energy
denominator approximation (CEDA) \cite{Gruning..Baerends.2002}.  We
compare these results with those of the BJ and related potentials as
well as with the AA-BJ ones obtained by applying the AASC$\alpha$
model to the BJ functional. It is shown that the AA-BJ results are
only slightly improved due to this application.  We also include some
previous AASC$\alpha$ model results obtained for the PW91 and GLLB
model functionals (in particular of the AA-PW91 and AA-m2 types, see
Ref. \cite{Karasiev-AA-02}).  These results show that all these models
are closer to EXX than the BJ or AA-BJ ones in the case of diatomic
molecules.

\section{Results and discussion}
 
  Table \ref{tab:table1} shows the Hartree-Fock total energies and the
  energy differences for KS-x-{\it only} methods calculated when the
  virial relation is used (except for the \lq\lq BJ(conv)"
  column). The energy differences of the approximate methods (eight
  last columns) should be compared to the exact exchange (EXX) values
  shown in the second column or to the KS(x-{\it only}) values obtained by
the iterative procedure described in Ref. \cite{DD} (shown in the third column),
which are very close to the EXX values.

All approximate methods shown in Table \ref{tab:table1}, except for
the BJ potential, are exact for the singlet state of a two-electron
system (H$_2$ molecule).  The orbital-dependent methods for the
response potential term (GLLB, AA-m2, CEDA and LHF) provide very good
approximations to the EXX energies, the CEDA and LHF values are only
~2 mHartrees higher than the corresponding EXX energies. The AA-PW91
potential, taken here as an example of an asymptotically-adjusted
potential where the response term is modeled by a conventional PW91
DFT functional was found to yield a good approximation to the EXX
energy; the largest difference is 39.4 mHartees for the F$_2$ molecule
(as compared to the EXX difference which is 8.6 mHartrees).

  The BJ differences still are significantly larger than those arising
  from all other approximate methods.  The negative value in Table
  \ref{tab:table1} shows that the corresponding energy is lower than
  the HF value, i.e. the variational principle is not satisfied.
  Large errors in the total energy also affect significantly the
  calculated atomization energies and the predicted equilibrium
  geometries.

  The ionization potentials approximated by the highest occupied
  molecular orbital (HOMO) energy are presented in Table
  \ref{tab:table2}. The situation is similar: the GLLB, AA-m2, CEDA
  and LHF methods provide an excellent approximation to the EXX
  values. The AA-PW91 values are also very close to the EXX for all
  molecules presented in Table except for the N$_2$ and FH.  The BJ
  potential underestimates the ionization potential by an amount of
  $\sim$30-50\%. The BJ potential is calculated without shift as
  suggested in \cite{Becke-Johnson-2006}. By applying the shift, the HOMO
  energies will be equal to the corresponding HF values (but not to the
  OEP ones).

\begin{center}
\begin{table*}[ht]
\hfill{}
\caption{\label{tab:table1} Full-numerical Hartree-Fock (HF) total
  energies (in a.u.)  and differences between KS-x-{\it only} and HF
  total energies (in mHartrees) calculated at the experimental
  geometries.
$^a$Values are taken from Ref. \cite{Ivanov-Hirata-Bartlett-1999}. 
$^b$From Ref. \cite{Karasiev-AA-02}. 
$^c$From Ref. \cite{Gruning..Baerends.2002} .
$^d$From Ref. \cite{Sala.Goerling.2003}.
}
\hfill{}
\begin{ruledtabular}
\begin{tabular}{lccccccccccc}
    & \tiny HF & \tiny EXX\tablenotemark[1] & \tiny KS(x-{\it only}) 
& \tiny AA-PW91\tablenotemark[2]
& \tiny GLLB\tablenotemark[2] 
& \tiny AA-m2\tablenotemark[2]  
& \tiny CEDA\tablenotemark[3]
& \tiny LHF\tablenotemark[4] & \tiny BJ(vir) & \tiny BJ(conv) & \tiny AA-BJ \\
\hline
\hspace*{10pt}\\ [-8pt]
H$_2$    & -1.1336   & 0.0 & 0.0 & 0.0  &  0.0 &  0.0 &0.0 & 0.0  & -80.6 & 0.8 & 0.0  \\
FH       & -100.0708 & 2.0 & 2.2 & 17.0 &  5.9 &  3.7 &    &      & 533.1 & 10.4 & 6.6 \\
N$_2$    & -108.9931 & 5.2 & 5.7 & 30.4 &  11.4&  8.3 & 7.7 & 7.3 & 239.4 & 9.9  & 9.5 \\
CO       & -112.7909 & 5.1 & 5.6 & 31.1 &  12.2&  9.1 & 7.6 & 7.2 & 323.0 & 12.6 & 11.5\\
F$_2$    & -198.7722 & 8.6 & 9.3 & 39.4 &  16.8&  13.6&     &     & 1124.9& 23.1 & 18.1\\
\end{tabular}
\end{ruledtabular}
\end{table*}
\end{center}

\begin{center}
\begin{table*}[ht]
\hfill{}
\caption{\label{tab:table2}
Ionization potential approximated by the negative of the HOMO energies (in eV). 
$^a$From Ref. \cite{Gorling-1999}. $^b$Ref. \cite{Karasiev-AA-02}.
$^c$Ref. \cite{Ivanov-Hirata-Bartlett-1999}. $^d$N$_2\rightarrow$N$^+_2(^2\Sigma_g)$.
$^e$N$_2\rightarrow$N$^+_2(^2\Pi_u)$.
$^f$From Ref. \cite{Gruning..Baerends.2002} .
$^g$From Ref. \cite{Sala-Goerling.2001}.
}
\hfill{}
\begin{ruledtabular}
\begin{tabular}{lccccccccc}
    & HF & EXX\tablenotemark[1] 
& AA-PW91\tablenotemark[2] 
&     GLLB\tablenotemark[2] 
&   AA-m2\tablenotemark[2]  
& CEDA\tablenotemark[6]
& LHF\tablenotemark[7]
& BJ  & AA-BJ\\
\hline
\hspace*{10pt}\\ [-8pt]
H$_2$    & 16.2 & 16.2 & 16.2 &  16.2 &  16.2 & 16.2 & 16.2 & 10.1 & 16.1\\
FH       & 17.7 & 17.4 & 13.3 &  16.5 &  18.2 &      &      &  9.4 & 11.0 \\
N$_2$\tablenotemark[4]  
         & 17.3 & 17.2\tablenotemark[3] 
                   &
                         12.7
                               &  15.5 &  17.0 & 17.1 & & 9.9 & 10.4  \\
N$_2$\tablenotemark[5]    
         & 16.7 & 18.1\tablenotemark[3]
                       & 13.9 &  16.2 &  18.0 & 18.5 & &10.7 & 11.2\\
CO       & 15.1 & 14.1 & 11.1 &  13.7 &  15.0 & 15.0 & 15.0  & 8.4 & 9.0 \\
F$_2$    & 18.2 & 14.5 & 13.4 &  15.9 &  18.7 &      &       & 9.3 & 11.1 \\
\end{tabular}
\end{ruledtabular}
\end{table*}
\end{center}

\begin{center}
\begin{table*}[ht]
\hfill{}
\caption{\label{tab:table3}
Bond lengths (in angstroms) obtained from different x-{\it only} methods.
}
\hfill{}
\begin{ruledtabular}
\begin{tabular}{lccccc}
    & HF & KS(x-{\it only}) & BJ(vir) & BJ(conv)  & AA-BJ\\
\hline
\hspace*{10pt}\\ [-8pt]
H$_2$    & 0.734 & 0.734 & --     & 0.732 & 0.734 \\
FH       & 0.897 & 0.896 & 0.860 & 0.895 & 0.896 \\
N$_2$    & 1.065 & 1.065 & 1.066 & 1.065 & 1.065 \\
CO       & 1.102 & 1.101 & 1.086 & 1.099 & 1.099\\
\end{tabular}
\end{ruledtabular}
\end{table*}
\end{center}

The asymptotically-adjusted, GLLB, CEDA and LHF exchange potentials, 
as well as the BJ model potential, provide an excellent
approximation to the exact exchange Kohn-Sham potential.  
The asymptotically-adjusted potentials based on the DFT approximation for the
response term (the AA-PW91) are invariant w.r.t. unitary
transformation of orbitals (as is also the BJ potential); however,
the AA-m2 and GLLB ones are not.  The advantages of the previously proposed
AA-PW91, AA-m2 and GLLB models (and of the CEDA and LHF methods) 
are the following: (i) For these models $E_{\rm
  vir}= E_{\rm conv}$ (notations from \cite{Staroverov08}); these models
constitute examples of situations where although $(E_{\rm vir}- E_{\rm
  conv})=0$, the potential is still only approximate; (ii) the energies
obtained  by the AA, GLLB models and by the CEDA and LHF methods 
satisfy the variational principle:
$E^{\rm HF}\leq E^{\rm OEP}\leq E_{\rm vir}^{\rm approx}$, which is not the
case for the BJ model; (iii) the virial energy for the AA and for the GLLB
models is an excellent approximation to the OEP/EXX energy. This is not the
case for the model potential of Refs. \cite{Staroverov08,Becke-Johnson-2006}, in spite
of the fact that the BJ model is in excellent agreement with
the OEP exchange potential; (iv) the ionization potential approximated by the
HOMO energy is a property entirely defined by the effective potential. The
recently proposed BJ model potential fails to adequately describe
this property, in contrast with the AA-PW91, GLLB and AA-m2 models and with
CEDA and LHF methods, where 
the agreement with the EXX values is, in most cases, excellent.

Some of the shortcomings of the model potentials from Refs.
\cite{Staroverov08,Becke-Johnson-2006} may be eliminated, however, by scaling the
model response term by the AASC$\alpha$ method, as it was done in Ref.
\cite{Karasiev-AA-02}.  A rigorous variational justification for this
type of scaling is given in Eqs. (12) through (15) of
Ref. \cite{Karasiev-AA-02}.  By applying this procedure to the BJ model, the new
AA-BJ potential (for the spin-unpolarized case) reads 
\be
v_{\rm x}^{\rm AA-BJ}=v_{\rm
  S}+\alpha_{\rm x}[\{\psi_i\}] \sqrt{\tau[\{\psi_i\}]/\rho}\, , 
\label{E1}
\ee
where $\tau$ is the kinetic energy density.
The self-consistent constant $\alpha_{\rm x}$ is defined by
\begin{equation}
\alpha_{\rm x}[\{\psi_i\}] = \frac{E_{\rm x}[\{\psi_i\}]-E_{\rm xLP}[v_S,\rho]}
                {E_{\rm xLP}[\sqrt{\tau[\{\psi_i\}]/\rho},\rho]}\, .
\label{M8}
\end{equation}
The energies for the new AA-BJ model
(we emphasize that $E_{\rm vir}= E_{\rm conv}$ for AA-BJ) are
presented in last column of Table I.  The AA-BJ energies are slightly
closer to the EXX values than $E_{\rm conv}$ values for the original
BJ model (BJ(conv) column in Table I). Moreover, modified AA-BJ model
eliminates the ambiguity with regard to the choice of functional for the
evaluation of the exchange energy (conventional or virial).  The
ionization potentials presented in Table II also are slightly improved
in the AA-BJ model as compared to the original BJ values.

In Table \ref{tab:table3} we present some bond lengths values obtained
from several different x-{\it only} methods.  We do not include the diatomic
molecule F$_2$ as it does not bind at the level of an x-{\it only}
approximation. The results show that for H$_2$, FH and N$_2$, the AA-BJ bond
lengths coincide up to three decimals with those obtained by means of
the exact Kohn-Sham x-{\it only} method.  For the case of CO, there is a
difference of 0.002 angstroms between the AA-BJ and the KS(x-{\it only})
result. The BJ(vir) geometries differ from the KS(x-{\it only}) results for 
all four diatomics presented in Table. The BJ(conv) bond lengths coincides with
the exact KS(x-{\it only}) results for the case of N$_2$, and differ for other molecules.

\section{Conclusions}

All potentials discussed here have the same structure: they comprise a
Slater potential plus a modeled response term. As a result, the
computational cost for all models is approximately the same, except
that for the AA-PW91, GLLB, AA-m2, CEDA, LHF cases, where an
additional exact-exchange energy term has to be calculated.  The same
term, however, must also be calculated for the BJ model potential when
the total energy is obtained not by means of the virial relation,
(where it is denoted as $E_{\rm vir}$ in Ref.  \cite{Staroverov08}),
but through the exact exchange term (in which case, it is denoted as
$E_{\rm conv}$).  The AASC$\alpha$ method is a simple procedure which
permits to transform the BJ model potential into a new AA-BJ potential
which satisfies the Levy-Perdew virial relation without increasing the
computational cost (in fact, this procedure may be
applied to model potentials of any structure without reducing
computational efficiency).  However, the calculated AA-BJ values show only a
slight improvement on the BJ ones, except for the bond lengths, which are in
excellent agreement with the KS x-only results.

\end{document}